\documentclass[aps,prb,showpacs,twocolumn,amsmath,amssymb,superscriptaddress]{revtex4}
\usepackage{graphicx}
\usepackage{Jonasmacros}
\usepackage{bm}

\begin{document}
\date{\today}
\title{Detection of exchange interaction in STM measurements through Fano-like interference effects}
\author{J. Fransson}
\email{jonasf@lanl.gov}
\affiliation{Theoretical Division, Los Alamos National Laboratory, Los Alamos, New Mexico 87545, USA}
\affiliation{Center for Nonlinear Studies, Los Alamos National Laboratory, Los Alamos, New Mexico 87545, USA}

\begin{abstract}
We address Fano-like interference effects in scanning tunneling microscopy (STM) measurements of nanoscale systems, e.g. two-level systems. Common for these systems is that second order tunneling contributions give rise to interference effects that cause suppressed transmission through the system for certain energies. The suppressed transmission is measurable either in the differential conductance or in the bias voltage derivative thereof. 
\end{abstract}
\maketitle

\section{Introduction}
\label{sec-introduction}
There are various techniques that allows one to detect and manipulate spin states in the solid state, which attract lots of interest. A partial list include optical detection of electron spin resonance (ESR) in a single molecule,\cite{koehler1993} tunneling through a quantum dot,\cite{engel2001} and, more recently, ESR-scanning tunneling microscopy (ESR-STM) technique.\cite{manassen1989,durkan2002} The interest in ESR-STM is due to the possibility of manipulating single spins,\cite{manoharan2002,balatsky2002,koppens2006} something which is crucial in spintronics and quantum information. Experimentally, modulation in the tunneling current has been observed by STM using spin-unpolarized electron beam.\cite{engel2001,manassen1989} Lately, there has also been a growing interest in using spin-polarized electron beam for direct detection of spin structures,\cite{wiesendanger2000} as well as utilizing the inelastic electron scanning tunneling spectroscopy (IETS) for detection of local spatial variations in electron-boson coupling in molecular systems.\cite{stipe1998,grobis2005}

Typically in STM measurements with an object located on a substrate surface, the tunneling current can either go directly between the STM tip and the substrate or go via the object. The tunneling electrons are thus branched between different pathways, which gives rise to interference effects when the partial waves merge into one in the tip or the substrate.\cite{madhaven2001} This interference leads to a suppressed transmission probability for the tunneling electrons at certain energies. The suppressed transmission is a fingerprint of Fano resonances,\cite{fano1961} and generally appear in systems where tunneling electrons are branched between different pathways. Recently, Fano resonances have been studied in double and triple quantum dots systems,\cite{kubala2002,guevara2003,lu2005} where the different pathways are constituted of the different quantum dots.

Fano resonances can be realized in a variety of system, ranging from systems with interactions between continuum states and a localized state, to systems where the branching of the wave function through diatomic molecules. The case originally discussed by Fano,\cite{fano1961} here reformulated for the purpose of STM and nanoscale systems, interference occur between the different tunneling paths in real space, one path going through the sample to the substrate whereas the other goes directly into the substrate. This idea is further exploited in works by Kubala and K\"onig,\cite{kubala2002} Ladr\'on de Guevara \etal.\cite{guevara2003}, and Lu \etal.\cite{lu2005} There are also many reports of IETS-STM measurements on molecules adsorbed onto metallic surfaces, which apart from the inelastic contributions also show Fanolike signatures in the transport data.\cite{bayman1981,hahn2000,smit2002,agrait2002,wang2004} Theoretically, these features can be interpreted as interference effects between the different tunneling channels.\cite{davis1970,persson1987,tikhodeev2001,galperin2004}

Here, we address the Fano interference effects that arise due to different pathways in phase space. In single level systems this phase space interference is can be caused by an effective spin-flip tunneling rate that is comparable to the spin-preserving tunneling rate.\cite{franssonSPFano2007} In two-level systems, tunneling paths such as $\ket{N=2,n}\rightarrow\ket{N=1,m}\rightarrow\ket{N=2,n'}\rightarrow\ket{N=1,m}\rightarrow\ket{N=2,n}$ give rise to phase space interference. Here, $N=1,2$, denote the number of electrons in the state, whereas $n,m$ are states indices.\cite{franssonSTsplit2007}  Although such tunneling paths are of second order, they provide significant contributions to the transmission coefficient, and hence to the conductance, which accounts for the interference effect discussed here. The interference effects are sufficiently described in mean-field approximation of the sample correlation functions, hence, we do not discuss any fluctuations caused by electronic correlations or by the couplings to the tip and substrate.

The paper is organized as follows. We introduce the formalism for the transport calculations in Sec. \ref{sec-current}. We discuss the Fano like interference effects and their implications on two-level systems in Secs. \ref{sec-tls}, and we summarize the paper in Sec. \ref{sec-summary}.

\section{Tunneling current}
\label{sec-current}
The STM system we have in mind can generally be described by the model Hamiltonian
\begin{equation}
\Hamil=\Hamil_\tip+\Hamil_\sub+\Hamil_\text{sample}+\Hamil_T,
\label{eq-STM}
\end{equation}
where the first and second terms describe the electronic states in the tip and substrate, respectively. Here, we assume flat band free electron like models for the states in the tip/substrate, and define $\Hamil_\tip=\sum_{p\sigma\in\tip}\dote{p\sigma}\cdagger{p}\cc{p}$ and $\Hamil_\sub=\sum_{q\sigma\in\sub}\dote{q\sigma}\cdagger{q}\cc{q}$, for the tip and substrate, respectively. Henceforth, we let the wave vectors $p\ (q)$ belong to the tip (substrate). Creation (annihilation) of an electron at the energy $\leade{p(q)}$ is denoted by $\cdagger{p(q)}\ (\cc{p(q)})$, and we let $\sigma=\up,\down$ denote the spin projection. The third term contains the information about the sample that is meant to be studied in the STM experiment, whereas $\Hamil_T$ describes the tunneling between the sample and the tip and the substrate. The tunneling Hamiltonian can in general be written as
\begin{equation}
\Hamil_T=\sum_{pn\sigma}v_{pn\sigma}\cdagger{p}\dc{n\sigma}
	+\sum_{qn\sigma}v_{qn\sigma}\cdagger{q}\dc{n\sigma}+H.c.,
\label{eq-HT}
\end{equation}
where $v_{p(q)n\sigma}$ is the tunneling rate between the tip (substrate) and the sample, whereas $\ddagger{n\sigma}\ (\dc{n\sigma})$ denotes creation (annihilation) of a spin $\sigma$ electron at the $n$th level in the sample.

The current is derived by standard methods, i.e. in the stationary regime we have $J=J_\tip=-e\ddtinline\av{N_\tip(t)}$, which gives\cite{meir1992,jauho1994}
\begin{equation}
J=\frac{e}{h}\int{\cal T}(\omega)[f_\tip(\omega)-f_\sub(\omega)]d\omega,
\label{eq-TJ}
\end{equation}
where the transmission coefficient
\begin{equation}
{\cal T}(\omega)=\tr\bfGamma^\tip\bfG^r(\omega)\bfGamma^\sub\bfG^a(\omega).
\label{eq-T}
\end{equation}
Here, $\bfG^{r(a)}$ is the retarded (advanced) Green function (GF) of the sample which are to be determined later in the paper. The coupling between the sample and the tip is denoted by $\bfGamma^\tip_{nm\sigma}=2\pi\sum_pv_{pn\sigma}^*v_{pm\sigma}\delta(\omega-\leade{p})$ while the coupling between the sample and the substrate is analogously defined by $\bfGamma^\sub_{nm\sigma}=2\pi\sum_qv_{qn\sigma}^*v_{qm\sigma}\delta(\omega-\leade{q})$. In the expression for the current, Eq. (\ref{eq-TJ}), we have also defined the Fermi function $f_{\tip(\sub)}(\omega)=f(\omega-\mu_{\tip(\sub)})$ at the tip chemical potential $\mu_{\tip(\sub)}$. For later reference we define the bias voltage $eV=\mu_\tip-\mu_\sub$, where we use $\mu_\tip=\mu_0+eV$ and $\mu_\sub=\mu_0$ with the equilibrium chemical potential $\mu_0$.

The objective of the present paper is to describe the expected features in the transport characteristics that are caused by the Fanolike interference effects when the tunneling electron wave functions are branched between different pathways. It is thus sufficient to provide a mean-field expression for the GF of the sample, in which case the present level of transport equations, Eqs. (\ref{eq-TJ}) and (\ref{eq-T}), are valid.

\section{Probing the two-level system}
\label{sec-tls}
Two level systems are the natural extension of single spins coupled to an environment of de-localized electrons. The most interesting physics in two-level systems is related to the lowest two-electron singlet state and the triplet states, and the exchange splitting between those states. Here, we will discuss an approach to measure the singlet-triplet exchange splitting parameter $J$ in two-level system through the presence of Fano-like interference effects.

We begin our study by considering a diatomic molecule constituted by two identical atoms. One may think of the atoms as quantum dots (QDs) and the molecule as a double QD. We assume that the atoms are coupled through Coulomb and exchange interactions $U$ and $J$, respectively. For simplicity, we assume infinite intra-level Coulomb interactions in order to avoid two electrons in one of the atoms, and we also assume that the tunneling between the atoms is negligible. This set of assumptions is not crucial for the effect we discuss, it merely permits a convenient framework for heuristic and qualitative studies of the approach. The approach is straight forwardly generalized, which is done in the end of this section.

The transport measurements are assumed to be performed by means of STM of a molecule located on a substrate surface. We therefore assume that the atoms couple equally strong to the surface, while the STM tip may couple asymmetrically to the atoms. We model the system by the Hamiltonian given in Eq. (\ref{eq-STM}) where tip and substrate Hamiltonians are given in Sec. \ref{sec-current}, while the Hamiltonian for the molecule (sample) is given by
\begin{eqnarray}
\Hamil_\text{sample}&=&\sum_{n\sigma}\dote{0}\ddagger{n\sigma}\dc{n\sigma}
	-2J\bfS_1\cdot\bfS_2
\nonumber\\&&
	+(U-J/2)(n_{1\up}+n_{1\down})(n_{2\up}+n_{2\down}),
\end{eqnarray}
along with the condition that there may not be more than one electron in each level $n=1,2$. Here, $\bfS_n=(1/2)\sum_{\sigma\sigma'}\ddagger{n\sigma}{\hat \sigma}_{\sigma\sigma'}\dc{n\sigma'}$ is a spin operator, where $\hat{\sigma}=(\sigma^x,\sigma^y,\sigma^z)$ is the vector of Pauli spin matrices. A complete model of the structure we consider may be given by Eq. (1) in Ref. \onlinecite{rasander2006}.

In the given model, the sample can be either in the empty state $\ket{0}$, have one electron in either of the bonding or anti-bonding states
\begin{subequations}
\label{eq-1e}
\begin{eqnarray}
\ket{1,1(2)}&=&\frac{\ddagger{1\up(\down)}-\ddagger{2\up(\down)}}{\sqrt{2}}\ket{0},
\\
\ket{1,3(4)}&=&\frac{\ddagger{1\up(\down)}+\ddagger{2\up(\down)}}{\sqrt{2}}\ket{0},
\end{eqnarray}
\end{subequations}
with corresponding energies $E_{1n}=\dote{0}$, respectively, or have two electrons in the singlet 
\begin{equation}
\ket{2,1}=\ket{S=0,m_z=0}=
	\frac{\ddagger{2\down}\ddagger{1\up}-\ddagger{2\up}\ddagger{1\down}}
		{\sqrt{2}}\ket{0}
\label{eq-singlet}
\end{equation}
with energy $E_{21}=2\dote{0}+U-J/2$, or triplet 
\begin{subequations}
\label{eq-triplet}
\begin{eqnarray}
\ket{2,2}&=&\ket{S=1,m_z=0}
	=\frac{\ddagger{2\down}\ddagger{1\up}+\ddagger{2\up}\ddagger{1\down}}
		{\sqrt{2}}\ket{0}, 
\\
\ket{2,3}&=&\ket{S=z,m_z=1}=\ddagger{2\up}\ddagger{1\up}\ket{0},
\\
\ket{2,4}&=&\ket{S=1,m_z=-1}=\ddagger{2\down}\ddagger{1\down}\ket{0}
\end{eqnarray}
\end{subequations}
configurations, with energies $E_{2n}=2\dote{0}+U+J/2$, $n=2,3,4$. Here we use the notation $\ket{N,n}$ where $N=0,1,2$ is the number of electrons, whereas $n$ is a state label.

The main effect that may be used for measurements of the singlet-triplet splitting $J$, arise due to phase space interference effects between the two-electron singlet and triplet states. This interference results in states that interacts only very weakly with the environment of de-localized electrons which, in turn, generate conductance suppression at biases that correspond to the singlet and triplet state energies. The conductance suppression is a direct response of the states that interact weakly with the surrounding electron bath.

Qualitatively, one can understand the appearance of the localized states that interact weakly with the electron bath by considering the following. Assume that there are two electrons in the molecule and that they are configured in the singlet state $\ket{2,1}=(\ddagger{2\down}\ddagger{1\up}-\ddagger{2\up}\ddagger{1\down})\ket{0}/\sqrt{2}$. Removal of a spin $\up$ electron from the molecule can be done by removing an electron from either of the atoms, i.e. through the process $(\dc{1\up}+\dc{2\up})\ket{2,1}=-(\ddagger{1\down}+\ddagger{2\down})\ket{0}/\sqrt{2}=-\ket{1,4}$. Removal of a spin $\up$ electron thus always puts the sample into the anti-bonding one-electron spin $\down$ state, which is orthogonal to the other one-electron states by construction. Analogously, removing a spin $\down$ electron from the molecule being in the singlet state necessarily leads to a transition to the anti-bonding one-electron spin $\up$ state $\ket{1,3}$. Clearly, the system cannot undergo first order transitions between bonding one-electron states $\ket{1,1(2)}$ and the singlet state. In this respect one may therefore regard the singlet state as being decoupled from the bonding one-electron states.

Likewise we find that the triplet states can be regarded as being decoupled from the anti-bonding one-electron states, since removal of an electron from any of the triplet state configurations results in a bonding one-electron state.

Having the sample described in terms of its eigenstates, it is beneficial to write the sample Hamiltonian in diagonal form as
\begin{equation}
\Hamil_\text{sample}=\sum_{Nn}E_{Nn}\ket{N,n}\bra{N,n}.
\end{equation}
The tunneling Hamiltonian $\Hamil_T$ is in the present context given by Eq. (\ref{eq-HT}), although we disregard spin-flip in the hybridization between the localized and de-localized electrons. Rewriting the operators $\dc{n\sigma}$ in the tunneling Hamiltonian in terms of the eigenstates of the sample we obtain
\begin{eqnarray}
\Hamil_T&=&\sum_{Nnm}\biggl(
	\sum_{p\sigma}v_{p\sigma Nnm}\cdagger{p}
	+\sum_{q\sigma}v_{q\sigma Nnm}\cdagger{q}\biggr)
\nonumber\\&&\times
	\ket{N,n}\bra{N+1,m}+H.c.
\end{eqnarray}
Here, the tunneling rates, c.f. Ref. \onlinecite{rasander2006},
\begin{equation}
v_{p(q)\sigma Nnm}\equiv\bra{N,n}
	(v_{p(q)1\sigma}\dc{1\sigma}+v_{p(q)2\sigma}\dc{2\sigma})\ket{N+1,m}
\end{equation}
also include the matrix elements for single electron transitions in the sample.

The sample GF in this system becomes a $20\times20$ matrix, which in general is not diagonal. This is because of higher order transitions that significantly contribute to the electronic structure and transport through the sample. The system can be simplified by assuming that only the diagonal processes like $\ket{N+1,m}\rightarrow\ket{N,n}\rightarrow\ket{N+1,m}$ and off-diagonal processes like $\ket{N+1,m}\rightarrow\ket{N,n} \rightarrow\ket{N+1,m'}\rightarrow\ket{N,n}\rightarrow\ket{N+1,m}$ contribute to the tunneling. In fact, the off-diagonal processes we include into our scheme is sufficient in order to described the phase space interference in the present system. This simplification leads to a system of ten $2\times2$ matrix equations, each of which is analytically solvable.

Consider the GFs\cite{franssonPRL2002,fransson2005}
\begin{equation}
G_{Nnmm'}(t,t')=\eqgr{\X{nm}{NN+1}(t)}{\X{m'n}{N+1N}(t')},
\end{equation}
where $\X{nm}{NN+1}\equiv\ket{N,n}\bra{N+1,m}$. The discussed simplifications leads to the general equations of motion
\begin{eqnarray}
\biggl(i\ddt-\Delta_{Nmn}\biggr)G_{Nnmm'}(t,t')=
	\delta(t-t')P_{Nnmm'}(t)
\nonumber\\
	+\sum_{\nu\mu}P_{Nnm\mu}(t)\kbint V_{Nn\mu\mu'n}(t,t'')
\nonumber\\\times
		G_{Nn\mu'm'}(t'',t')dt'',
\label{eq-SCGF}
\end{eqnarray}
the transition energy $\Delta_{Nmn}=E_{N+1m}-E_{Nn}$, the end-factors $P_{Nnmm'}(t)=\av{\anticom{\X{nm}{NN+1}}{\X{m'n}{N+1N}}(t)}$, whereas 
\begin{eqnarray}
V_{Nn\mu\mu'n}(t,t')&=&
	\sum_{p\sigma}v_{p\sigma Nn\mu}^*v_{p\sigma Nn\mu'}g_{p\sigma}(t,t')
\nonumber\\&&
	+\sum_{q\sigma}v_{q\sigma Nn\mu}^*v_{q\sigma Nn\mu'}g_{q\sigma}(t,t').
\end{eqnarray}
Here, $g_{p(q)\sigma}(t,t')$ is the GF for a free electron in the tip (substrate) satisfying the equation $(i\ddtinline-\leade{p(q)})g_{p(q)\sigma}(t,t')=\delta(t-t')$.

The GF in Eq. (\ref{eq-SCGF}) should be self-consistently solved for each value in the parameter space of the Hamiltonian Eq. (\ref{eq-STM}), bias voltage and temperature, according to procedure lined out in Ref. \onlinecite{fransson2005}. Performing the calculations is such a way would provide a non-equilibrium description of the interference effects discussed here. For simplicity and in order to focus on the physical mechanism and effect we omit such a treatment. The self-consistent calculations are expected to change the quantitative results, however, the qualitative features will remain the same as in the present study. Here we discuss the physics without the self-consistency condition by further simplifying the equation of motion. Without losing information of the phase space interference effects, we assume that the end-factors $P_{Nnmm'}=\delta_{mm'}$. In this we assume that the off-diagonal occupation numbers are negligible, however, the off-diagonal GFs are non-vanishing and, furthermore, they provide important contributions to the widths of the localized states in the sample. Moreover, we focus on the transitions between the one- and two-electron states. In absence of spin-flip processes the equations for transitions between the one-electron spin $\up$ and $\down$ states are equal. It is hence sufficient to omit any reference to the spin degree of freedom, and therefore we consider processes that couple the two-electron singlet and states $\ket{2,1}$ and $\ket{2,n}$, $n=2,3,4$, through a one-electron state. 

It is worth pointing out that Eq. (\ref{eq-SCGF}) is mean-field description of the sample, thus neglecting any type of fluctuations caused by the couplings between the sample and the electron baths in the tip and substrate. Fluctuations caused by the couplings to the tip and substrate electrons could by taken into to consideration by making use of the diagrammatic technique described in Refs. \cite{franssonPRL2002,fransson2005}. Especially interesting would be to also consider the singlet-triplet Kondo effect recently discussed by Paaske \etal\cite{paaske2006} in the present context. However, a discussion of the Kondo effect and other correlation effects is beyond the scope of the present paper. Here, we focus on the possibility of measuring the singlet-triplet splitting $J$ through direct transport measurements, which is sufficiently described within the present mean-field theory.

Transitions between the different triplet state configurations do not give rise to the interference effects we discuss in this paper. Therefore, it is sufficient to study the coupling between the singlet state and one of the triplet states at the time. It is, moreover, sufficient to consider only the interference between the singlet and, say, the triplet state $\ket{2,2}$. The interference arising between the singlet and the other triplet configurations merely renormalizes the coefficients in the final expression for the transmission.

The Fourier transformed retarded GF for transitions between the two-electron states $\ket{2,1}$ and $\ket{2,2}$ through the one-electron state $n$ is then given by
\begin{equation}
\bfG_n^r(\omega)=\frac{
	\left(\begin{array}{cc}
		\omega-\Delta_{2n}+\frac{i}{2}\Gamma_{n2} &
			-\frac{i}{2}\Gamma_{1n2} \\
			-\frac{i}{2}\Gamma_{2n1} &
		\omega-\Delta_{1n}+\frac{i}{2}\Gamma_{n1}
	\end{array}\right)}{C_n(\omega)},
\end{equation}
where
\begin{eqnarray}
\Gamma_{mnm'}&\equiv&-2\im V_{1nmm'n}^r(\omega)
\nonumber\\
	&=&2\pi\sum_{k\sigma}v_{k\sigma1nm}^*v_{k\sigma1nm'}\delta(\omega-\leade{k}),
\end{eqnarray}
and $\Gamma_{nm}\equiv\Gamma_{mnm}$, defines the combined coupling between the sample and the tip and the substrate. Further, we define the coupling to the tip (substrate) by $\Gamma^{\tip(\sub)}_{mnm'}=2\pi\sum_{p(q)\sigma}v_{p(q)\sigma1nm}^*v_{p(q)\sigma1nm'}\delta(\omega-\leade{p(q)})$, such that $\Gamma_{mnm'}=\Gamma_{mnm'}^\tip+\Gamma_{mnm'}^\sub$. For a shorter notation we have also put $\Delta_{mn}\equiv\Delta_{1mn}$. Finally, the denominator $C_n(\omega)=(\omega-\omega_{n+})(\omega-\omega_{n-})$ contains the poles $\omega_{n\pm}$ of the GF.

\subsection{Symmetric coupling to the substrate}
\label{ssec-sym}
Now, we assume that the atoms in the diatomic molecule couple equally strong to the substrate, i.e. $v_{qn\sigma}=v_{q\sigma}$. Then, the coupling matrices for the tunneling between the sample and the substrate take the simple forms
\begin{subequations}
\label{eq-Gsub}
\begin{eqnarray}
\bfGamma^\sub_{\ket{1,1}}&=&\Gamma_0^\sub
	\left(\begin{array}{cc} 0 & 0 \\ 0 & 1 \end{array}\right)
\\
\bfGamma^\sub_{\ket{1,3}}&=&\Gamma_0^\sub
	\left(\begin{array}{cc} 1 & 0 \\ 0 & 0 \end{array}\right)
\end{eqnarray}
\end{subequations}
We further assume that the tunneling rate between the tip and the molecule can be parametrized by $v_{pn\sigma}=\gamma_nv_{p\sigma}$. The coupling matrices for the tunneling between the tip and the sample then become
\begin{subequations}
\label{eq-Gtip}
\begin{eqnarray}
\bfGamma^\tip_{\ket{1,1}}&=&\frac{\Gamma_0^\tip}{4}
	\left(\begin{array}{cc} (\gamma_1-\gamma_2)^2 & -\gamma_1^2+\gamma_2^2 \\
		-\gamma_1^2+\gamma_2^2 & (\gamma_1+\gamma_2)^2 \end{array}\right)
\label{eq-Gtipb}\\
\bfGamma^\tip_{\ket{1,3}}&=&\frac{\Gamma_0^\tip}{4}
	\left(\begin{array}{cc} (\gamma_1+\gamma_2)^2 & -\gamma_1^2+\gamma_2^2 \\
		-\gamma_1^2+\gamma_2^2 & (\gamma_1-\gamma_2)^2 \end{array}\right)
\label{eq-Gtipab}
\end{eqnarray}
\end{subequations}
Finally, we assume that $\Gamma_0^\tip=\lambda\Gamma_0/2$ and $\Gamma_0^\sub=\Gamma_0/2$, where $\lambda\ll1$, such that the broadening of the localized states $\bfGamma_{\ket{2,m}}\simeq\bfGamma_{\ket{2,m}}^\sub$. This final assumption is made in order to simplify the analytical treatment and yields the GF poles $\omega_{1\pm}$ for transitions through the bonding state
\begin{equation}
\omega_{1\pm}=\dote{0}+U\mp\frac{J}{2}-\frac{i}{8}\Gamma_0(1\mp1),
\end{equation}
and the poles $\omega_{3\pm}$ for transition through the anti-bonding state
\begin{equation}
\omega_{3\pm}=\dote{0}+U\mp\frac{J}{2}-\frac{i}{8}\Gamma_0(1\pm1),
\end{equation}
since $\Delta_{m1}=\Delta_{m3}$, $m=1,\ldots,4$, and since the transition energies $\Delta_{11}=E_{21}-E_{11}=2\dote{0}+U-J/2-\dote{0}=\dote{0}+U-J/2$ and $\Delta_{21}=E_{22}-E_{11}=2\dote{0}+U+J/2-\dote{0}=\dote{0}+U+J/2$, which are associated with the singlet and triplet states, respectively. The poles of the GFs are thus located at the singlet and triplet energies, as expected. Moreover, from these expressions it is legible that the poles $\omega_{1+}$ and $\omega_{3-}$ acquire vanishing widths, which marks energies for which the corresponding transitions occur with vanishing probability. Hence, in the DOS of the sample there are states that interacts only weakly with the surrounding de-localized electrons. In this sense, there both broad and sharp peaks expected to appear at energies that are associated with both the singlet and triplet states.

The sharply peaked state in the sample is a result of the interference between tunneling electron waves that are branched in phase space. This branching is responsible for the coupling of the singlet and triplet states through the one-electron states. The interference that arise between the branches of the tunneling electron waves leads to both constructive and destructive interference which manifest itself through the widths of the localized states in the sample. In terms of this argument, the states that correspond to the poles $\omega_{1+}$ and $\omega_{3-}$ are subject to destructive interference which is sufficiently strong to remove any significant broadening of the state, it hence interact weakly with the surrounding electron bath(s). These states should display sharp peaks in the local DOS of the sample. The states that are associated with the poles $\omega_{1-}$ and $\omega_{3+}$ are subject to constructive interference which tend to increase the broadening of these states. Those states can, thus, be viewed as being strongly interacting with the environment.

\begin{figure}[t]
\begin{center}
\includegraphics[width=8.5cm]{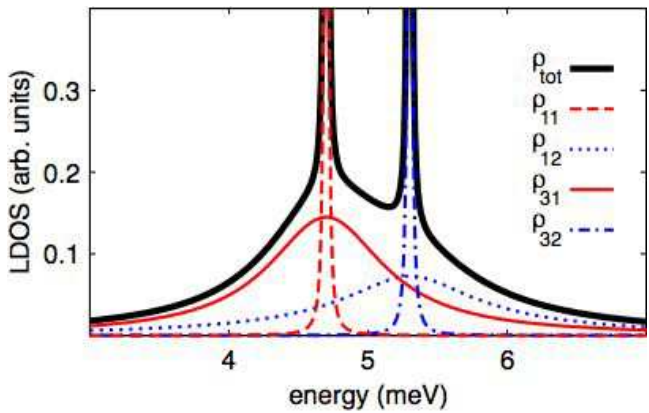}
\end{center}
\caption{(Color online) Typical DOS (bold) around the singlet and triplet states in the sample. The DOS associated with the poles $\omega_{1\pm}$ (dashed, dotted) and $\omega_{3\pm}$ (solid, dash-dotted) are plotted separately. Here we have the following parameters, $\dote{0}=2$, $U=3$, $J=0.6$, $\Gamma_0=2$ (units: meV), $\lambda=0.005$, $\gamma_1=0$, and $\gamma_2=1$.}
\label{fig-STdos}
\end{figure}
The plot in Fig. \ref{fig-STdos} (bold) illustrate a typical example of the calculated local DOS 
\begin{equation}
\rho_\text{tot}(\omega)=\sum_{nm}\rho_{nm}(\omega)
	=-\frac{1}{\pi}\sum_{nm}\im G^r_{1nmm}(\omega)
\end{equation}
for energies around the singlet and triplet in the present context. There are two major broad densities which peak at the singlet (solid) and triplet (dotted) energies, respectively. These states, which have a strong interaction with the environment, mediate the tunneling current through the molecule. On top of the broad densities there are narrow peaks centered at the singlet (dashed) and triplet (dash-dotted) energies, respectively. These are the states that interact weakly with the de-localized electrons. In the calculations these poles have a finite width due to the small coupling to the electrons in the tip, c.f. the results for the poles given in Ref. \onlinecite{franssonSTsplit2007}. The plots corresponds to the situation where the STM tip couples strongly to one of the atoms and only very weakly to the second.

The states in the local DOS which interact weakly with the tunneling electrons are the source of dips in the differential conductance through the system. This can be directly seen in the transmission coefficient ${\cal T}(\omega)$ for the system. Using the above assumptions and that the widths of the levels in the sample can be written as $\Gamma_{ij}=\Gamma^\tip_{ij}+\delta_{ij}\delta_{i2}\Gamma^\sub_2$, we arrive at the transmission coefficient
\begin{eqnarray}
{\cal T}(\omega)&=&
	\lambda\biggl(\frac{\Gamma_0}{4}\biggr)^2(\gamma_1+\gamma_2)^2
	\biggl(
	\biggl|\frac{\omega-\dote{0}-U+J/2}{C_1(\omega)}\biggr|^2
\nonumber\\&&
	+\biggl|\frac{\omega-\dote{0}-U-J/2}{C_3(\omega)}\biggr|^2
	\biggr)
\end{eqnarray}
for the transmission through the two-electron states. This expression clearly shows that the transmission dips at the energies associated with the singlet and triplet states, and that the distance between the dips equals $J$, i.e. the exchange splitting parameter.

\begin{figure}[t]
\begin{center}
\includegraphics[width=8.5cm]{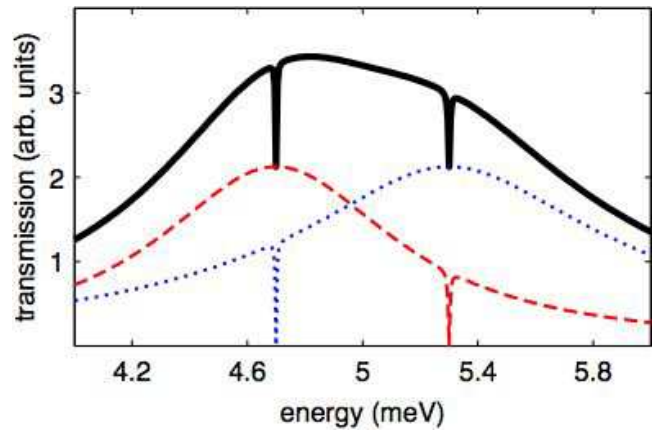}
\end{center}
\caption{(Color online) Transmission coefficient for the system with local DOS plotted in Fig. \ref{fig-STdos}. Total (bold) and partial transmissions through the channel which couples the two-electron state via the bonding (dotted) and anti-bonding (dashed) one-electron states.}
\label{fig-STtrans}
\end{figure}

The total transmission coefficient for the tunneling around the singlet and triplet states is shown in Fig. \ref{fig-STtrans} (bold), along with the partial transmission coefficients for through channel that couples the two-electron states via the bonding (dotted) and anti-bonding (dashed) one-electron states. The plot demonstrate that the partial transmissions have an anti-resonance on the energy that corresponds to the singlet and triplet states, respectively. These anti-resonances are reflected in the total transmission.

Whenever the broadening of the quantum levels in the sample is larger than the singlet-triplet splitting $J$, there will not be two distinct peaks associated with the singlet and triplet state in the differential conductance. In addition, visibility of the dips in the differential conductance require low temperatures since the transmission has to be convoluted with the thermal distribution functions, c.f. Eq. (\ref{eq-TJ}). Therefore, in order to resolve the singlet-triplet splitting under those circumstances, one should preferably measure $d^2I/dV^2$ rather than the differential conductance. Those measurements provide further information about the long-lived states in the sample in terms of very sharp features at the energies corresponding to those states.

\begin{figure}[t]
\begin{center}
\includegraphics[width=8.5cm]{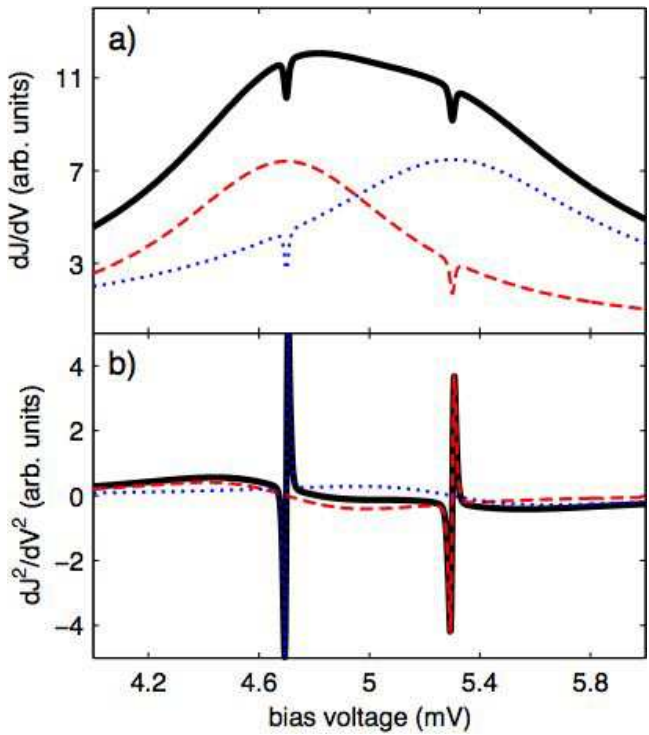}
\end{center}
\caption{(Color online) Transport characteristics for the system with local DOS given plotted in Fig. \ref{fig-STdos}. a) Total differential conductance (bold) and the partial differential conductance through the channel which couples the two-electron states via the bonding (dotted) and anti-bonding (dashed) states. b) Total and partial bias voltage derivatives $d^2I/dV^2$. Here, $T=0.05$ K, while other parameters as in Fig. \ref{fig-STdos}.}
\label{fig-STdJ}
\end{figure}
In Fig. \ref{fig-STdJ} we plot the calculated transport characteristics corresponding to the DOS in Fig. \ref{fig-STdos}. The upper panel, Fig. \ref{fig-STdJ} a), shows the total differential conductance $dI/dV$ (bold) and the partial differential conductance for the tunneling through the channel which couples the two-electron states via the bonding (dotted) and anti-bonding (dashed) one-electron states, respectively. The broad electron densities give rise to wide conductance peaks at the singlet and triplet states which, however, are not clearly distinguishable in the total $dI/dV$ since the level broadening is larger than the singlet-triplet exchange splitting. Nonetheless, there are noticeable features, or dips, in the $dI/dV$ at the energies for the singlet and triplet states. These dips are signatures of the states that are weakly interacting with the tunneling electrons. The large level broadening smears out the differential conductance and makes a unambiguous read-out of the singlet and triplet states difficult. However, second derivative of the current $d^2I/dV^2$ will provide the information that is required for an unambiguous read-out since the conductance dips will display sharp features at the positions for the singlet and triplet states, as can be seen in Fig. \ref{fig-STdJ} b). This panel shows the total (bold) and partial $d^2I/dV^2$ for the tunneling through the channel which couples the two-electron states via the bonding (dotted) and anti-bonding (dashed) one-electron states. 

\subsection{Asymmetric coupling to the substrate}
\label{ssec-asym}
Generally, the atoms in the molecule may couple asymmetrically to the substrate. The sample may, on the other hand, be comprised of a single entity with two levels. In both these cases we can assume that the hybridization can be parametrized according to $v_{qn\sigma}=\kappa_nv_{q\sigma}$.

The STM tip was already in the previous subsection assumed to couple asymmetrically to the sample, hence, we merely have to re-derive the poles for the GFs and the transmission coefficients using the additional parametrization. The coupling matrices $\bfGamma^\sub_{\ket{2,n}}$ take the form of the ones given in Eq. (\ref{eq-Gtip}), however, with $\gamma_n$ replaced by $\kappa_n$. We still assume that the tunneling between the atoms in the molecule is negligible. Then, the localized states will be found at the energies
\begin{subequations}
\begin{eqnarray}
\omega_{1\pm}&=&\dote{0}+U-\frac{i}{16}\Gamma_0(\kappa_1^2+\kappa_2^2)
\\&&
	\pm\frac{1}{2}\sqrt{\biggl(J-\frac{i}{4}\kappa_1\kappa_2\Gamma_0\biggr)^2
		-(\kappa_1^2-\kappa_2^2)^2\biggl(\frac{\Gamma_0}{8}\biggr)^2}
\nonumber\\
\omega_{3\pm}&=&\dote{0}+U-\frac{i}{16}\Gamma_0(\kappa_1^2+\kappa_2^2)
\\&&
	\pm\frac{1}{2}\sqrt{\biggl(J+\frac{i}{4}\kappa_1\kappa_2\Gamma_0\biggr)^2
		-(\kappa_1^2-\kappa_2^2)^2\biggl(\frac{\Gamma_0}{8}\biggr)^2}
\nonumber
\end{eqnarray}
\end{subequations}
for the transitions through the bonding and anti-bonding states, respectively. Again we assume that the level broadening is dominated by the coupling between the sample and the substrate surface. The expressions for the poles show that both solutions $\omega_{n\pm}$, for both $n=1,3$, acquire an appreciable width for asymmetric coupling between the sample levels and the substrate. Thus, for sufficiently large asymmetry between the couplings, there will not be any states present that interacts weakly with the surrounding electron bath(s), and give rise to the conductance dips. This is an indication that large asymmetric coupling between the levels in the sample and the substrate is detrimental to the ability of observing the sharp features in the $dI/dV$ or $d^2I/dV^2$. Presence of such features will also depend on the asymmetry in the coupling between the sample levels and the STM tip. In general, in order to obtain localized states that interact weakly with the surrounding electron bath(s), the levels in the sample should be fairly symmetrically coupled either the tip or the substrate. Good symmetry in the coupling to both the tip and the sample enhances the amplitude of the sharp features in the $d^2I/dV^2$.

The transmission through the sample that is mediated by the two-electron states is modified according to
\begin{eqnarray}
{\cal T}(\omega)&=&\lambda\biggl(\frac{\Gamma_0}{4}\biggr)^2
	(\gamma_1\kappa_1+\gamma_2\kappa_2)^2	
	\biggl(\biggl|\frac{\omega-\dote{0}-U+qJ/2}{C_1(\omega)}\biggr|^2
\nonumber\\&&
	+\biggl|\frac{\omega-\dote{0}-U-qJ/2}{C_3(\omega)}\biggr|^2\biggr),
\end{eqnarray}
where the asymmetry of the coupling introduces the factor
\begin{equation}
q=\frac{\gamma_1\kappa_2+\gamma_2\kappa_1}
	{\gamma_1\kappa_1+\gamma_2\kappa_2}.
\end{equation}
Hence, apart from affecting the overall pre-factor, asymmetric coupling between of the levels in the sample and the substrate shifts the positions of the transmission dips. Quantitatively, coupling asymmetries $\gamma_2/\gamma_1\gtrsim0.7$ and $\kappa_2/\kappa_1\gtrsim0.7$ yields about 6\% shift of the transmission dips. Hence, even for asymmetric coupling this will be a reasonable good measurement of $J$.

\subsection{Non-resonant levels}
\label{ssec-nrl}
The case thus far considered applies to system where the atoms are equivalent, so that the levels in the two atoms are resonant with one another. Generally, the two levels do not need to be resonant | they may also be spin-split, in which case the first term in the sample Hamiltonian is modified according to
\begin{equation}
\sum_{n\sigma}\dote{n\sigma}\ddagger{n\sigma}\dc{n\sigma},
\end{equation}
where $\dote{n\sigma}=\dote{n}-\sigma\Delta_n/2$. Here, $\Delta_n$ is the spin-split of the $n$th level which is imposed by either internal and/or external magnetic fields.

The singlet and triplet states remain the same as they are given Eqs. (\ref{eq-singlet}) and (\ref{eq-triplet}) while the energies for the respective state may change. The singlet state energy becomes $E_{21}=\dote{1}+\dote{2}+U-J/2$, whereas the triplet state energies become $E_{22}=\dote{1}+\dote{2}+U+J/2$, $E_{23}=E_{22}-(\Delta_1+\Delta_2)/2$, and $E_{24}=E_{22}+(\Delta_1+\Delta_2)/2$. The one-electron states is in this modified model preferably described by
\begin{eqnarray}
\ket{1,n}=\{\delta_{n1}\ddagger{1\up}+\delta_{n2}\ddagger{2\down}
	+\delta_{n3}\ddagger{2\up}+\delta_{n4}\ddagger{2\down}\}\ket{0},
\end{eqnarray}
with corresponding energies $E_{1n}=(\delta_{n1}+\delta_{n3})\dote{n\up}+(\delta_{n2}+\delta_{n4})\dote{n\down}$.

We consider the general case in the sense that we relax the assumption of the levels being symmetrically coupled to the substrate. We thus assume that the tunneling rates can be parametrized according to $v_{pn\sigma}=\gamma_nv_{p\sigma}$ and $v_{qn\sigma}=\kappa_nv_{q\sigma}$, as in the previous subsection. The coupling matrices between the electrons in the STM tip and in the sample become
\begin{subequations}
\begin{eqnarray}
\bfGamma^\tip_{\ket{1,1}}&=&\frac{\Gamma^\tip_0}{2}\gamma_2^2
	\left(\begin{array}{cc} 1 & 1 \\ 1 & 1 \end{array}\right),
\\
\bfGamma^\tip_{\ket{1,3}}&=&\frac{\Gamma^\tip_0}{2}\gamma_1^2
	\left(\begin{array}{cc} 1 & -1 \\ -1 & 1 \end{array}\right),
\end{eqnarray}
\end{subequations}
while the couplings matrices between electrons in the substrate and the sample are obtained by the replacements $\gamma_n\rightarrow\kappa_n$ and $\tip\rightarrow\sub$. Here, we have only considered the coupling matrices for the singlet state and the triplet state configuration $\ket{2,2}=\ket{S=1,m_z=0}$. The coupling matrices that involve the other triplet configurations are the same.

In the same approximation as previously, i.e. assuming that the broadening of the levels in the sample is dominated by the hybridization with the electrons in the substrate, we obtain the following poles to the GFs which involve the the two-electron states $\ket{2,1}=\ket{S=0,m_z=0}$ and $\ket{2,2}$
\begin{subequations}
\begin{eqnarray}
\omega_{1\pm}&=&=\frac{\dote{1\up}+\dote{2\up}+2U}{2}
	-\frac{i}{4}\kappa_2^2\Gamma_0
\nonumber\\&&
		\pm\frac{1}{2}\sqrt{J^2-(\kappa_2^2\Gamma_0/2)^2}
\\
\omega_{3\pm}&=&=\frac{\dote{1\up}+\dote{2\up}+2U}{2}
	-\frac{i}{4}\kappa_1^2\Gamma_0
\nonumber\\&&
		\pm\frac{1}{2}\sqrt{J^2-(\kappa_1^2\Gamma_0/2)^2}.
\end{eqnarray}
\end{subequations}
The poles $\omega_{2\pm}$ and $\omega_{4\pm}$, for the spin $\down$ channels, are obtained by letting $\up\rightarrow\down$ in the equations above. From the form of the poles we deduce that, in the spin-degenerate case there may be sharp localized state only in the case when $J\ll\kappa_n\Gamma_0/2$. In the symmetric case it should be possible to measure $J$ in cases where the singlet-triplet splitting is small. Then, the theory presented in the previous subsections apply, and therefore we proceed our discussion for the spin-dependent case.

The spin-dependent case we consider is based on the assumption $\dote{1}=\dote{2}=\dote{0}$ and that the level spin-split is uniform, i.e. $\Delta_1=\Delta_2=\Delta_0$. We also assume that the levels in the sample couple symmetrically to the substrate. In this case we can use one-electron states given in Eq. (\ref{eq-1e}) and the coupling matrices given in Eqs. (\ref{eq-Gsub}) and (\ref{eq-Gtip}). Considering spin-dependent transport enable studies of the spin-splitting of the two-electron triplet states. The singlet state couples to all triplet configurations, through the one-electron state, which therefore will display sharp localized state at different energies. These different energies correspond to the spin-splitting of the triplet configurations. Due to the spin-dependence, there appear four sets of poles which couple the singlet with each of the triplet state through the one-electron states. For transitions to the one-electron spin $\up$ state we have
\begin{subequations}
\begin{eqnarray}
\omega_{1\pm}^{\ket{2,2}}&=&\dote{0}+\frac{\Delta_0}{2}
	+U\mp\frac{J}{2}-\frac{i}{8}\Gamma_0(1\mp1),
\\
\omega_{3\pm}^{\ket{2,2}}&=&\dote{0}+\frac{\Delta_0}{2}
	+U\mp\frac{J}{2}-\frac{i}{8}\Gamma_0(1\pm1),
\\
\omega_{1\pm}^{\ket{2,3}}&=&\omega_{1\pm}^{\ket{2,2}}-\Delta_0(1\mp1),
\\
\omega_{3\pm}^{\ket{2,3}}&=&\omega_{3\pm}^{\ket{2,2}}-\Delta_0(1\mp1),
\end{eqnarray}
\end{subequations}
while the poles associated with transitions to the one-electron spin $\down$ states are given by
\begin{subequations}
\begin{eqnarray}
\omega_{2\pm}^{\ket{2,2}}&=&\dote{0}-\frac{\Delta_0}{2}
	+U\mp\frac{J}{2}-\frac{i}{8}\Gamma_0(1\mp1),
\\
\omega_{4\pm}^{\ket{2,2}}&=&\dote{0}-\frac{\Delta_0}{2}
	+U\mp\frac{J}{2}-\frac{i}{8}\Gamma_0(1\pm1),
\\
\omega_{2\pm}^{\ket{2,4}}&=&\omega_{2\pm}^{\ket{2,2}}+\Delta_0(1\mp1),
\\
\omega_{4\pm}^{\ket{2,4}}&=&\omega_{4\pm}^{\ket{2,2}}+\Delta_0(1\mp1),
\end{eqnarray}
\end{subequations}

Note that the poles are equal in pairs according to $\omega_{1\pm}^{\ket{2,2(3)}}=\omega_{2\pm}^{\ket{2,4(2)}}$ and $\omega_{3\pm}^{\ket{2,2(3)}}=\omega_{4\pm}^{\ket{2,4(2)}}$. Due to these equalities we expect four sharp peaks in the local DOS of the sample which are associated with the two-electron states, two for the singlet and two for the triplet states. The reason is clear since the one-electron spin $\up$ states do not at all couple to the triplet state $\ket{2,4}=\ddagger{2\down}\ddagger{1\down}\ket{0}$. Therefore, there cannot appear any sharp peak associated with the corresponding transitions. The same argument applies to the spin $\down$ channel. In any other respect, each spin channel can be treated separately by means of the theory developed in Secs. \ref{ssec-sym} and \ref{ssec-asym}.

\section{Summary}
\label{sec-summary}
We have presented a theoretical prediction of the possibility to measure the singlet-triplet exchange interaction parameter $J$ through Fano-like interference effects. We argue for measurements of $d^2I/dV^2$ for extraction of $J$, since the level broadening in many realistic situations is larger than the singlet-triplet splitting. The level broadening therefore effectively smears out possible identification features in the differential conductance which makes the read-out of $J$ difficult. We also address the issue about asymmetric coupling between the sample levels and the substrate, and find that moderate asymmetries preserve a reasonable good measurability of $J$. Finally, we address the question of non-resonant levels. There will not be any Fano-like interference effects between which enable singlet-triplet splitting read-out unless the levels are resonant. However, uniform spin-splitting of the levels do not destroy the measurability of the singlet-triplet read-out. In the spin-split case we predict four sharp features in the $d^2I/dV^2$ instead of two, which enable pairwise read-out of i) the spin-split and ii) the singlet-triplet splitting. Experimental results on two-level systems using the addressed set-up would be very intriguing and may open novel approaches to information storage.

\acknowledgements
This work has been supported by US DOE, LDRD and BES, and was carried out under the auspices of the NNSA of the US DOE at LANL under Contract No. DE-AC52-06NA25396. The author thanks A. V. Balatsky for useful discussions.

\end{document}